\documentclass[11pt]{article}
\usepackage[utf8x]{inputenc}
\usepackage[english]{babel}
\usepackage{microtype}
\usepackage{enumitem}
\usepackage{multicol}
\usepackage{threeparttable}
\usepackage{lmodern}
\usepackage{amsfonts,amsmath,amsthm,amssymb}
\usepackage{datetime}

\usepackage[separate-uncertainty]{siunitx}
\usepackage{tikz}
\usepackage{graphicx}
\usepackage{float}
\usepackage{subcaption}
\usepackage{pdflscape}
\graphicspath{{img/}}
\DeclareGraphicsExtensions{.pdf,.png,.jpg}
\usepackage{parskip}
\usepackage[final]{showkeys}
\usepackage{amsmath}
\usepackage{graphicx}
\usepackage{feynmp}
\makeatletter
\def\endfmffile{
  \fmfcmd{\p@rcent\space the end.^^J end.^^J endinput;}
  \if@fmfio
    \immediate\closeout\@outfmf
  \fi
  \ifnum\pdfshellescape>\z@
    \immediate\write18{mpost \thefmffile}
  \fi}
\makeatother

\mathchardef\mhyphen="2D

\newdateformat{monthyeardate}{%
	\monthname[\THEMONTH] \THEYEAR}
	
\newcommand\rurl[1]{\href{https://arxiv.org/abs/#1}{\nolinkurl{#1}}}

\usepackage[
 a4paper,
 total={10cm,23.7cm},          
 top=30mm, left=30mm,        
 headsep=10mm,               
 footskip=10mm               
]{geometry}

\usepackage[
 pdftex,                        
 bookmarks=true,                
 bookmarksdepth=3,              
 colorlinks=true,               
 linkcolor=black,               
 citecolor=black,               
 urlcolor=black,                 
]{hyperref}

\usepackage[english, capitalise]{cleveref}
\usepackage{hyperref}
\usepackage{authblk}

\begin{document}

\title{Reconstruction of $\tau$ lepton pair invariant mass using an artificial neural network}

\author[a]{P. B\"artschi\footnote{Corresponding author: pascal.baertschi@uzh.ch (P. B\"artschi)}}
\author[a]{C. Galloni}
\author[b]{C. Lange}
\author[a]{B. Kilminster}
\affil[a]{Physik Institut, Universit\"at Z\"urich, CH-8057 Z\"urich, Switzerland}
\affil[b]{CERN, CH-1211 Geneva, Switzerland}
\date{}

\maketitle

\begin{abstract}
The reconstruction of the invariant mass of $\tau$ lepton pairs is important for analyses containing Higgs and Z bosons decaying to $\tau^{+}\tau^{-}$, but highly challenging due to the neutrinos from the $\tau$ lepton decays, which cannot be measured in the detector. In this paper, we demonstrate how artificial neural networks can be used to reconstruct the mass of a di-$\tau$ system and compare this procedure to an algorithm used by the CMS Collaboration for this purpose. We find that the neural network output shows a smaller bias and better resolution of the di-$\tau$ mass reconstruction and an improved discrimination between a Higgs boson signal and the Drell-Yan background with a much shorter computation time.
\end{abstract}

\section{Introduction}
\label{sec:intro}
Tau leptons can be produced singly or in pairs through the decay of heavier mesons and baryons, or through the decay of Standard Model bosons that are created in particle collisions. The $Z^{0}/\gamma^{*}$ and Higgs (H) bosons are the only particles that can mediate resonant di-$\tau$ production, and Higgs boson production has been recently observed in this decay by the CMS and ATLAS experiment at CERN \cite{cmshiggstotautau}\cite{atlashiggstotautau}. 
The reconstruction of the di-$\tau$ system invariant mass is fundamental in distinguishing between the Z and the Higgs boson. The dominant background in identifying the $\mathrm{H} \rightarrow \tau^{-}\tau^{+}$ signal is the Drell-Yan (DY) process of di-$\tau$ production, therefore, it is important to distinguish between the two processes. However, the reconstruction of the mass of the di-$\tau$ system is challenging. The $\tau$ lepton decays after a short time into leptons or hadrons, both accompanied by neutrinos. Charged leptons and hadrons can be observed in the detector, thus they are called visible particles in this paper. Neutrinos, on the other hand, interact very weakly with matter and escape the experiment undetected, but can be identified in a relatively hermetic detector as used in the CMS and ATLAS experiments \cite{cmsdetector,atlasdetector} as an imbalance in the measured momentum calculated in the transverse plane. The missing transverse momentum ($\mathrm{MET}$) is defined as the negative vectorial sum of momenta in the transverse plane of all the visible particles produced in the collision. The momentum of the neutrinos in the beam direction, on the other hand, cannot be quantified, because it depends on the momentum of the quarks and gluons within the proton before the collision, which is only known statistically and not in any individual event.\\
The CMS Collaboration makes use of two different SVfit algorithms for the reconstruction of the di-$\tau$ mass. One is based on a likelihood method to reconstruct the mass on an event-by-event basis \cite{SVfit}, while the second, improved algorithm employs a likelihood function of arbitrary normalization \cite{SVfit_updated}. The latter algorithm is additionally able to reconstruct the kinematic properties of the di-$\tau$ system. The ATLAS and CDF Collaborations use a Missing Mass Calculation (MMC) method that is based on minimizing a likelihood function in the kinematically allowed detector phase space \cite{mmc}. The method presented in this paper is based on an artificial neural network (called neural network or NN from now on) to reconstruct the mass of the di-$\tau$ system. The neural network is implemented with the Python deep learning library "Keras" \cite{keras} and can be trained with a data set of simulated di-$\tau$ events, which contain all the known parameters of the visible particles and missing energy from neutrinos as input, and the mass of the simulated di-$\tau$ system as a training target. After the training, the neural network is able to calculate an approximate value of the mass of the di-$\tau$ system using the known parameters of the visible particles and the missing energy for any event with two $\tau$ leptons.\par\null\par \noindent
The paper is structured as follows: in section \ref{sec:simulated_events}, events with simulated $\tau$-lepton pairs are described, and are used to train and test the neural network. Section \ref{sec:nn_model} focuses on the specific configuration of the neural network used to deliver the best predictions in comparison with the true mass of the simulated di-$\tau$ system. The performance of the neural network is compared to a standard tool for the di-$\tau$ mass reconstruction used by the CMS Collaboration in section \ref{sec:results}. Conclusions are presented in the final section.

\section{Simulation and selection of \texorpdfstring{$\tau$}{tau} lepton pair events}
\label{sec:simulated_events}
\begin{figure}
	\begin{center}
		\includegraphics[scale= 0.5,keepaspectratio]{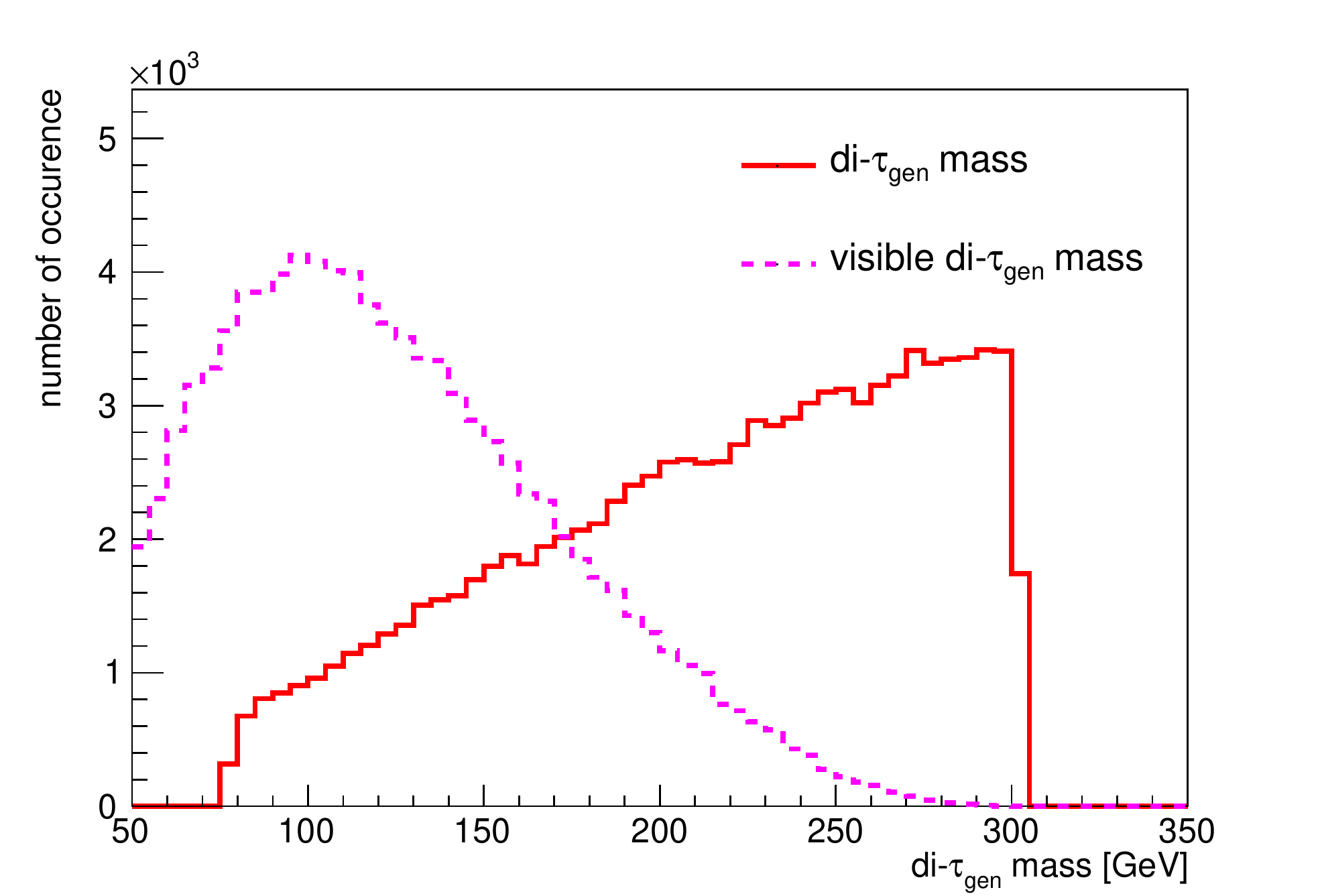} 
		\caption{Mass distribution of the generated events with applied cuts.}
		\label{fig:higgsmassvismass}
	\end{center}
\end{figure}

Monte Carlo simulation is used to generate $\tau^{+}\tau^{-}$ events at different masses. The events are generated using \verb|MadGraph_aMC@NLO 2.5.5| \cite{madgraph}, and then showered using \verb|PYTHIA 8.226| \cite{pythia_6.4}\cite{pythia_8.2}. The detector response is modeled using a simplified fast simulation (\verb|DELPHES 3.4.1| \cite{delphes}) of the phase-0 CMS detector with the acceptance and expected performance of the detector \cite{cmsdetector}. No additional proton-proton collisions (pile-up) are simulated.\\
DY events occur when a mediator $Z^{0}/\gamma^{*}$ is produced through quark-antiquark annihilation. This mediator can then decay into a pair of $\tau$ leptons.\\
In proton collisions, the main production processes for the Higgs boson, ordered from largest cross section to smallest, are gluon-gluon fusion, vector boson fusion, W and Z associated production, and t associated production. Events are simulated using the most dominant production mode via gluon-gluon fusion, which occurs through an intermediate heavy-quark loop that is dominated by the top quark \cite{higgs}. In order to increase the mass range over which the NN can reconstruct the di-$\tau$ system, the Higgs boson mass is artificially varied from 80 to 300 GeV in 5 GeV steps. The fully leptonic, semi-leptonic and fully hadronic Higgs decay channels are included. \\
Events with produced Higgs bosons must pass selection requirements similar to those applied in the CMS $\mathrm{H} \rightarrow \tau^{+}\tau^{-}$ search \cite{SVfit_updated}. The $\mathrm{MET}$ must be greater than 20 GeV for all events, in order to make sure that the neutrino momenta are not pointing in opposite directions. For events where both $\tau$ leptons decay leptonically, the $e/\mu$ with the higher $p_{\mathrm{T}}$ must satisfy $p_{\mathrm{T}} > 20$ GeV and $|\eta| < 2.4$, with the other $e/\mu$ satisfying $p_{\mathrm{T}} > 10$ GeV and $|\eta| < 2.4$. For events where only one $\tau$ lepton decays leptonically, the $e/\mu$ must satisfy $p_{\mathrm{T}} > 20$ GeV and $|\eta| < 2.1$, with the vectorial sum of the visible decay products of the hadronically decaying $\tau$ lepton having $p_{\mathrm{T}} > 30$ GeV and $|\eta| < 2.3$. For the case of events with two $\tau$ leptons decaying hadronically, both vectorial sums must have $p_{\mathrm{T}} > 20$ GeV and $|\eta| < 2.1$.\\
Figure \ref{fig:higgsmassvismass} shows the generated mass and visible mass spectrum for the simulated Higgs boson decaying to $\tau$ leptons. The mass corresponds to the true value of the mediator, while the visible mass is obtained by summing the momenta of the final state decay particles, not including the neutrinos. The di-$\tau_{\mathrm{gen}}$ mass spectrum is not flat due to selection requirements that are related to detector acceptance and identification effects. The visible mass of the di-$\tau$ system (visible di-$\tau_{\mathrm{gen}}$ mass) shows a much different spectrum. In order to separate signal from background, it is desirable to achieve a di-$\tau$ mass as close to the generated value as possible, and so information about the neutrino products must be inferred to improve the mass resolution.
To prevent the neural network from learning the mass distribution of the training sample instead of the relationship between the inputs and the training target, the training sample consists of events from all the different Higgs boson masses, with the same number of events for each mass value.\\
\\
The neural network is trained with 270'000 events and the performance of the neural network is tested with 100'000 independent events. The inputs for the neural network are the following parameters: 4 binary numbers classifying the decay channel (fully leptonic, semi-leptonic or fully hadronic), the $p_\mathrm{T}$, $\eta$, $\phi$, energy, and invariant mass of the visible decay products from each $\tau$ lepton, the $\mathrm{MET}$ and $\phi$ of the $\mathrm{MET}$ vector, and the collinear di-$\tau$ mass. The collinear di-$\tau$ mass is the mass of the di-$\tau$ system computed assuming that the vectorial sum of the visible $\tau$ lepton decay products point along the original $\tau$ lepton direction. The collinear di-$\tau$ mass can be calculated as shown in equation \ref{eq:ditaumass_collinear} \cite{collinear}:

\begin{equation}
m_{\mathrm{coll}}=\frac{m_{di\mhyphen\tau_{\mathrm{vis}}}}{\sqrt{\frac{p_{\mathrm{T}_{\tau_{\mathrm{vis1}}}}}{p_{\mathrm{T}_{\tau_{\mathrm{vis1}}}}+\mathrm{MET_{1}}}\cdot\frac{p_{\mathrm{T}_{\tau_{\mathrm{vis2}}}}}{p_{\mathrm{T}_{\tau_{\mathrm{vis2}}}}+\mathrm{MET_{2}}}}} \:,
\label{eq:ditaumass_collinear}
\end{equation}

where $m_{\mathrm{coll}}$ stands for the collinear di-$\tau$ mass, $m_{di-\tau_{\mathrm{vis}}}$ for the visible di-$\tau$ mass, $p_{\mathrm{T}_{\tau_{\mathrm{vis1,2}}}}$ and $\mathrm{MET_{1,2}}$ for the visible transverse momentum and missing transverse momentum of $\tau_{1}$ and $\tau_{2}$, respectively.
\section{Model of the neural network}
\label{sec:nn_model}
The neural network used is feedforward, has fully connected layers, and is written with the Python deep learning library, Keras \cite{keras}. Keras is a set of high-level building blocks that implement the deep learning model, and it interfaces with a backend, which handles operations such as tensor products and convolutions. The backends used in this work are Theano \cite{theano} for running on CPU and Tensorflow \cite{tensorflow} for running on GPU. The model used for reconstructing the Higgs boson mass from the di-$\tau$ system is shown in Table \ref{tab:higgs_model}.
\begin{table}[ht]
	\centering
    \caption{Neural network model for reconstructing the Higgs boson mass from the di-$\tau$ system.}
	\begin{tabular}{lcc}
    	\\
		Layer & Number of neurons & Activation function\\
        \hline
		Input layer & 17 neurons & -\\
		1. hidden layer & 200 neurons & ReLU\\
		2. hidden layer & 200 neurons & ReLU\\
		3. hidden layer & 200 neurons & ReLU\\
		4. hidden layer & 200 neurons & ReLU\\
		Output layer & 1 neuron & ReLU\\
	\end{tabular}%
	\label{tab:higgs_model}
\end{table}
\\
Batch sizes of 128 and 400 epochs are used for the training. The activation function used is the commonly employed "rectified linear unit (ReLU)" and the loss function is the mean squared error (MSE), which is described by Eq. \ref{eq:MSE}:

\begin{equation}
C(w,b) = \frac{1}{n}\sum\limits_{i=1}^{n}(\hat{Y}_i(w_i,b_i)-Y_i)^{2}\:,
\label{eq:MSE}
\end{equation}
\\
where $\hat{Y}_{i}$ is the prediction, $Y_{i}$ is the training target value, $n$ is the number of the training events, $w_{i}$ are the weights and $b_{i}$ are the biases used for the prediction $\hat{Y}_{i}$.
The optimizer used is called \verb+adam+ which stands for Adaptive Moment Estimation \cite{adam}. While the choice of network structure has been evaluated carefully and seems rather optimal for the problem and data set under investigation, the structure would have to be adjusted in case of adding or removing variables. 
\\
Once the neural network is trained, the performance can be tested with an independent second sample of simulated events. A good quantifier for the performance of the neural network is the mean squared error of the deviation between the predicted value and the true value of the mass from the simulated di-$\tau$ system of the test sample, as well as the reconstructed di-$\tau$ mass resolution.

\section{Results}
\label{sec:results}
\subsection{SVfit}
The performance of the neural network is evaluated with respect to other current working methods for the reconstruction of the di-$\tau$ mass. For example the CMS Collaboration uses SVfit \cite{SVfit}, which reconstructs the mass of the di-$\tau$ system using dynamical likelihood techniques. The term dynamical likelihood techniques refers to likelihood-based methods used for the reconstruction of kinematic quantities on an event-by-event basis. The inputs to SVfit are the visible decay products of the $\tau$ leptons, $\mathrm{METx}$ and $\mathrm{METy}$ as well as the $\mathrm{MET}$ covariance matrix. The $\mathrm{MET}$ covariance matrix represents the expected resolution of the $\mathrm{MET}$ reconstruction in the detector (since the $\mathrm{MET}$ measurement is affected by the accuracy of the energy calibrations).

\subsection{Comparison between neural network and SVfit}
\label{subsec:higgsresultsstand}
\begin{figure}[ht]
	\begin{center}
		\includegraphics[scale= 0.6,keepaspectratio]{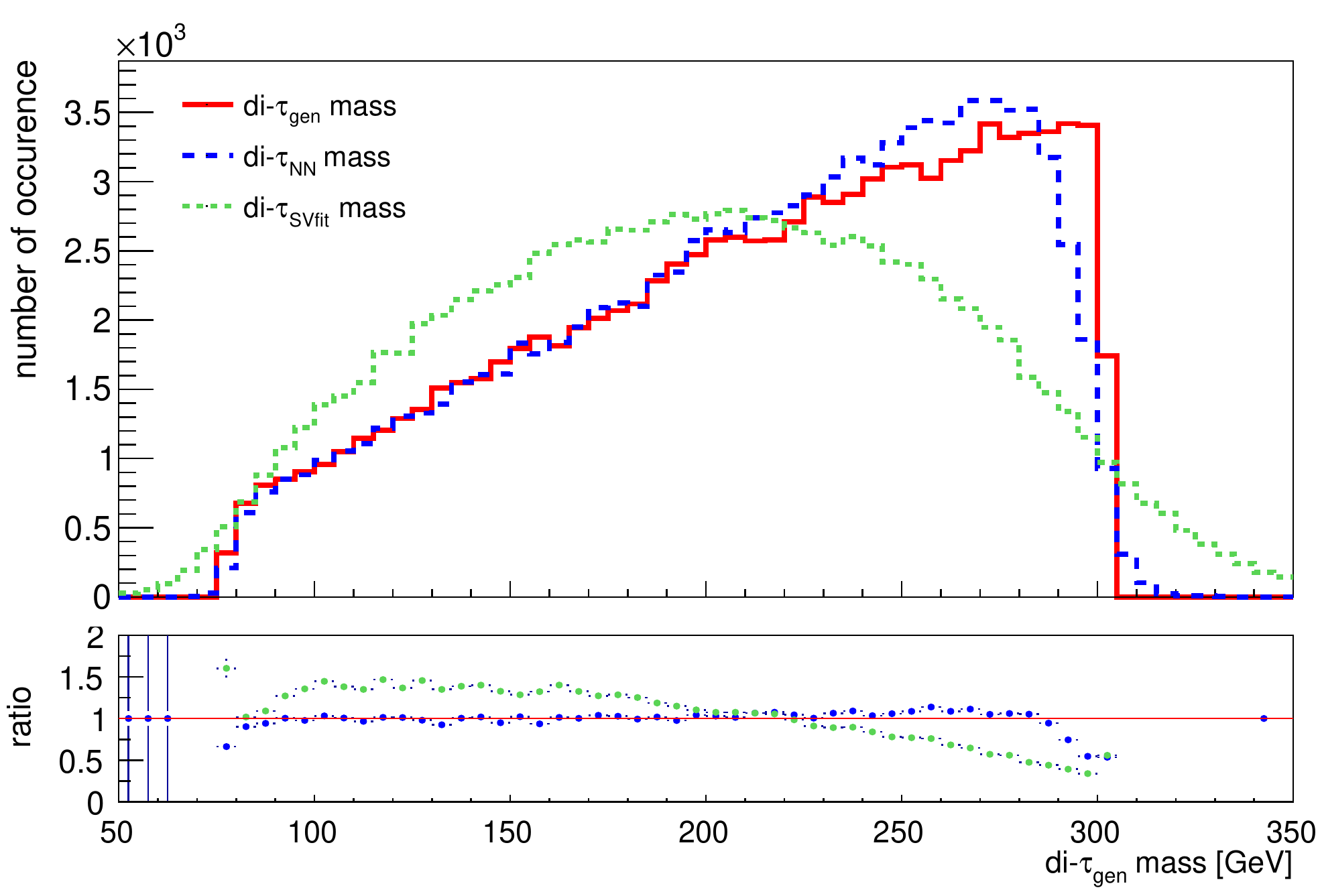} 
		\caption{Reconstructed mass of di-$\tau$ system. Di-$\tau_{gen}$ refers to the simulated mass values, di-$\tau_{NN}$ and di-$\tau_{SVfit}$ to the reconstructed mass values using the neural network and SVfit, respectively.}
		\label{fig:nnsvfit}
	\end{center}
\end{figure}

\begin{figure}[ht]
	\begin{center}
		\includegraphics[scale= 0.5,keepaspectratio]{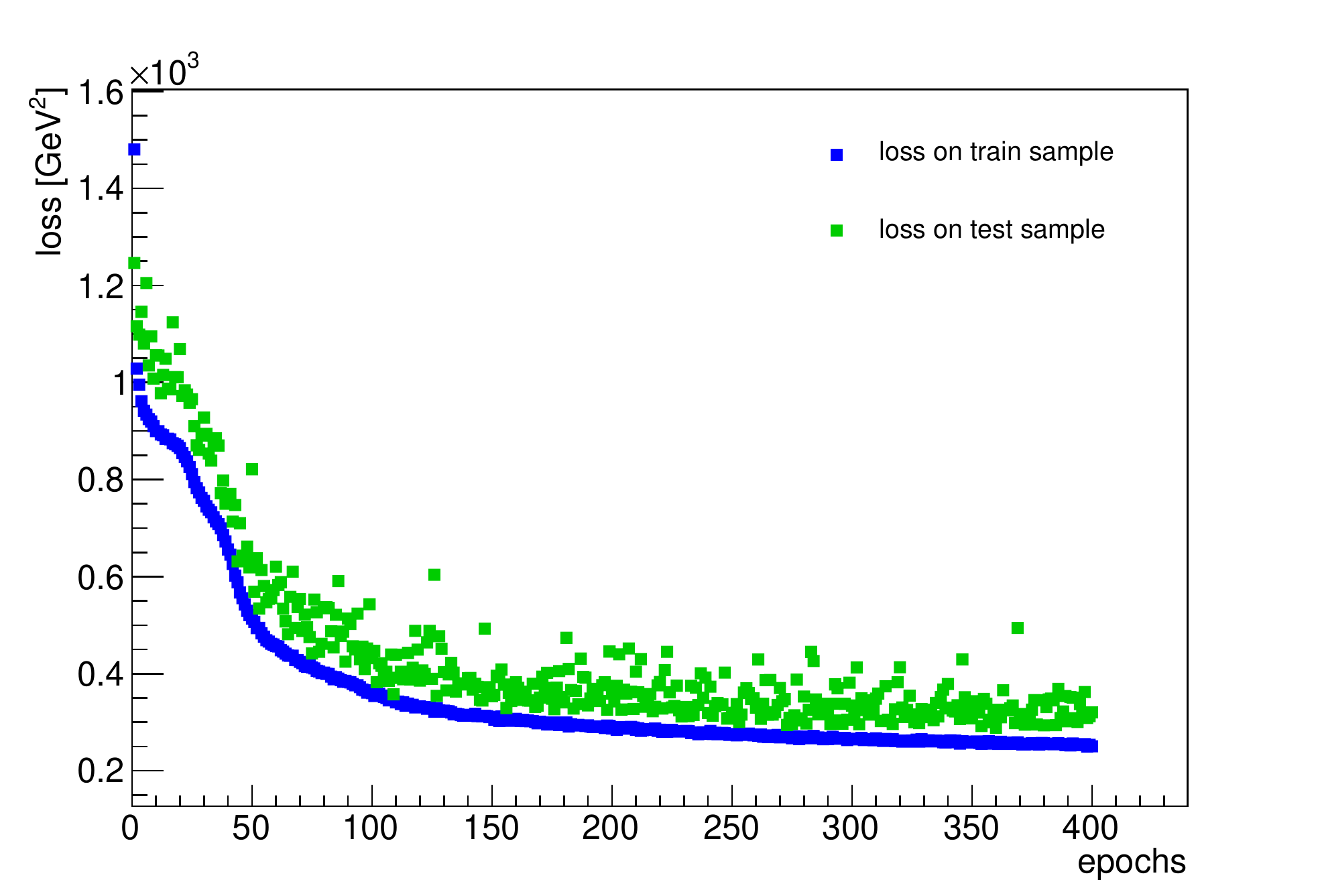} 
		\caption{The loss, as determined by the MSE, for the training and testing samples of the neural network.}
		\label{fig:nnsvfit_loss}
	\end{center}
\end{figure}

If not specified differently, events with fully leptonic, semi-leptonic and fully hadronic decays are used.
The predictions of the neural network are shown in figure \ref{fig:nnsvfit} in comparison with the results from SVfit and with the mass of the simulated events (di-$\tau_{\mathrm{gen}}$ mass). The shape of the di-$\tau_{\mathrm{gen}}$ mass distribution originates from the applied cuts. The predictions of the neural network show deviations at the limits of the mass range. Because there are more events in the higher mass range, this effect can most significantly be seen in the range of 250 GeV to 300 GeV. The deviations at the mass range limits are most probably caused by the distribution of events in the training sample, which does not contain any events below 80 GeV or above 300 GeV. However, the results from the neural network show less of a deviation than that of SVfit. Such a deviation in the mean of the reconstructed di-$\tau$ mass would not lead to a wrong result in an analysis as the same function would be applied to both data and simulation.
\\
In figure \ref{fig:nnsvfit_loss}, the loss (mean squared error) on the training and on the test sample can be seen. The loss shows no sign of overtraining. Overtraining would lead to a discrepancy between the loss on the training and on the test sample and is caused by the neural network describing minor fluctuations in the training sample instead of the underlying relationship.

\begin{figure}[ht]
	\begin{center}
		\includegraphics[scale= 0.6,keepaspectratio]{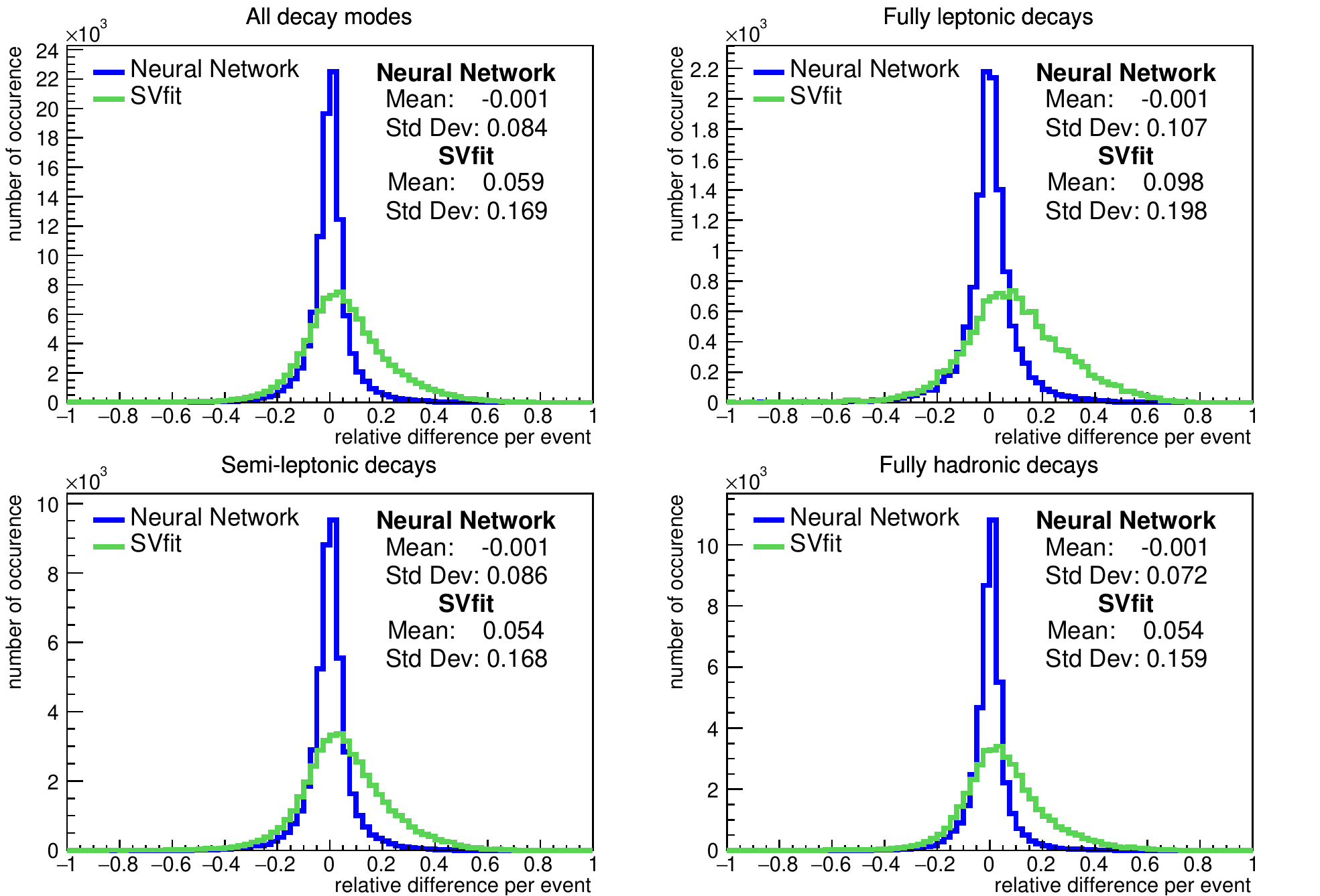} 
		\caption{Relative differences per event, as defined in the text, between the generated mass and the calculated mass determined from both the neural network and SVfit approaches.}
		\label{fig:nnsvfit_rescompar}
	\end{center}
\end{figure}

\begin{figure}[ht]
	\begin{center}
		\includegraphics[scale= 0.5,keepaspectratio]{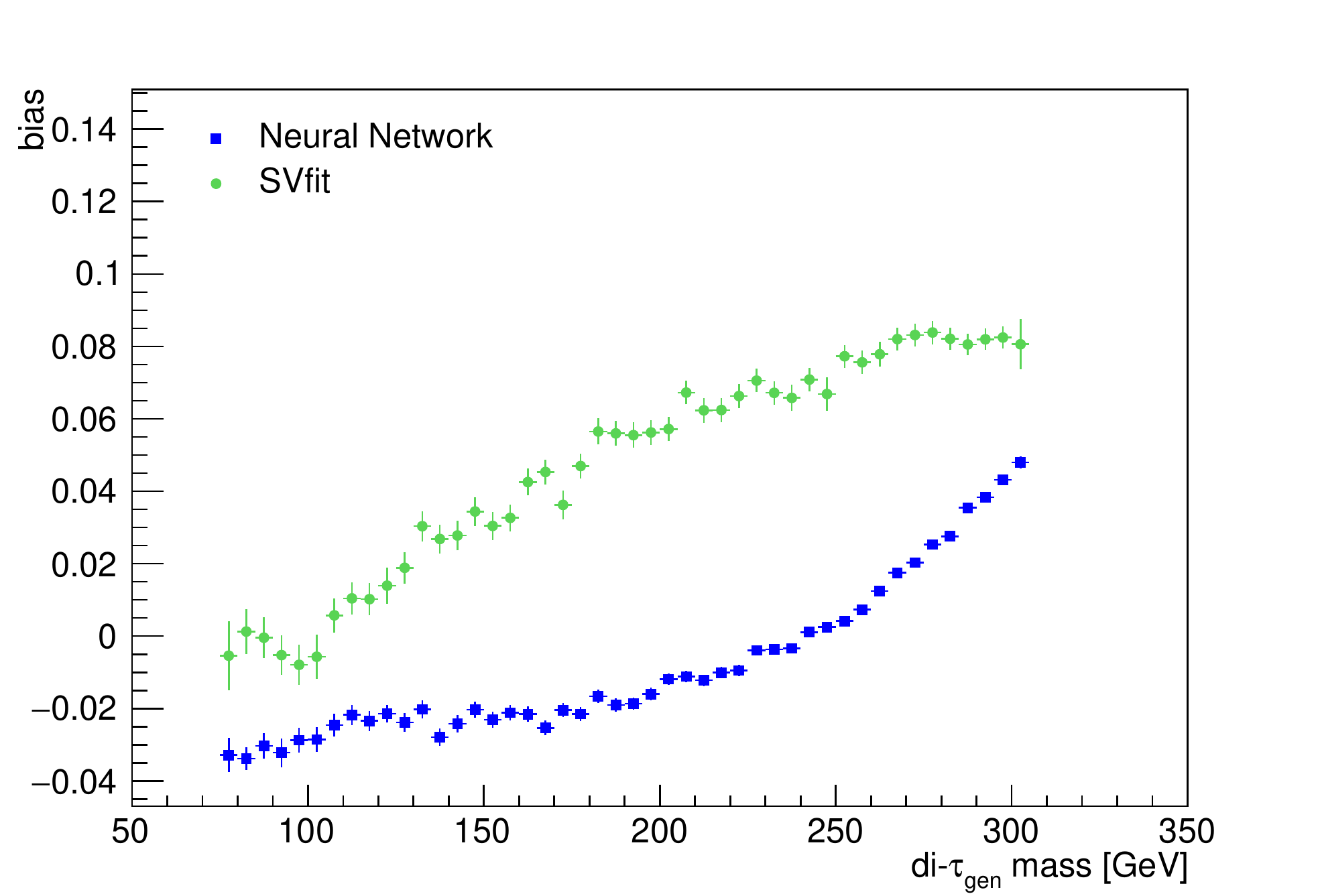} 
		\caption{Bias in the di-$\tau$ mass reconstruction per di-$\tau_{gen}$ mass.}
		\label{fig:nnsvfit_res_bias}
	\end{center}
\end{figure}

\begin{figure}[ht]
	\begin{center}
		\includegraphics[scale= 0.5,keepaspectratio]{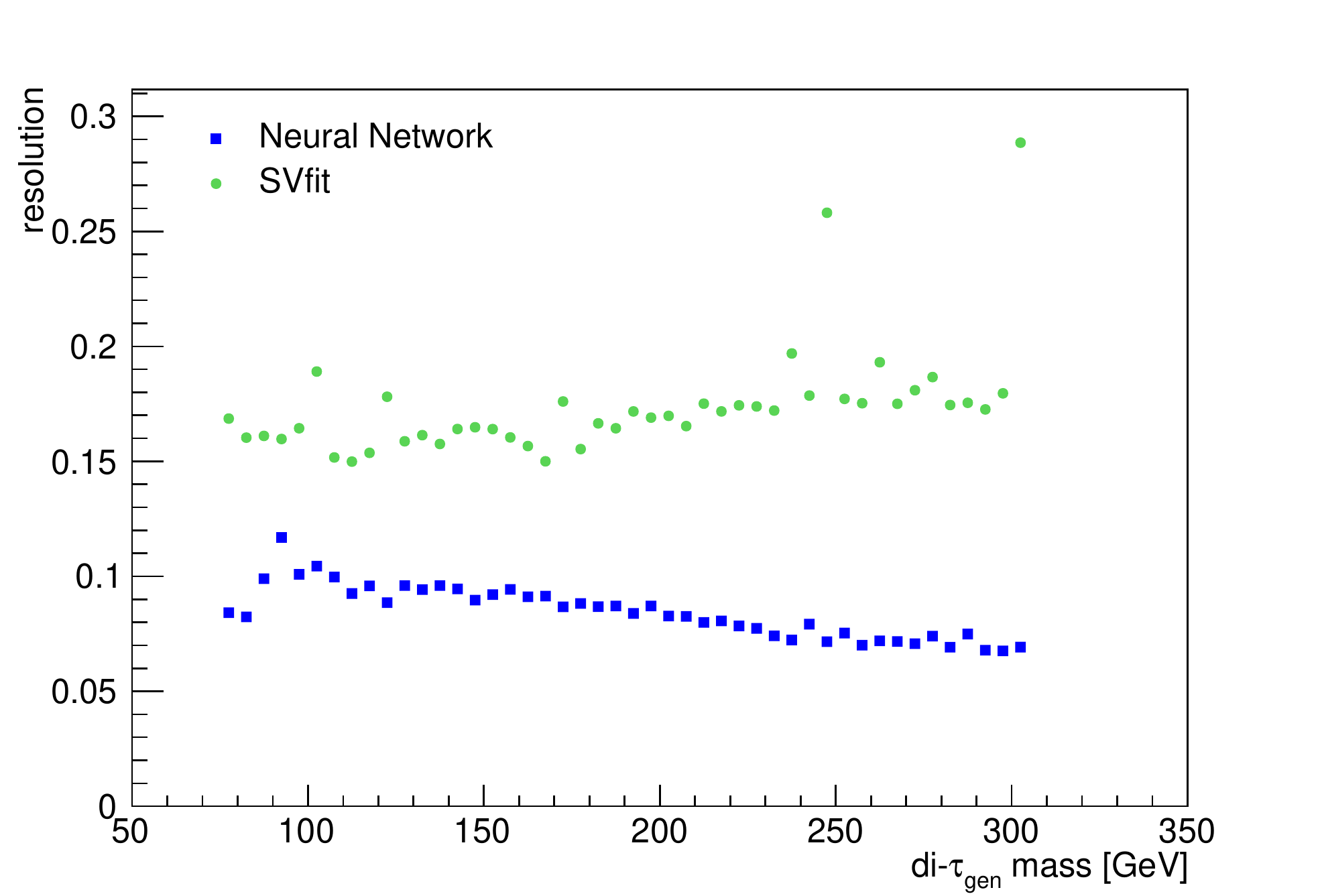} 
		\caption{The resolution of the di-$\tau$ mass reconstruction per di-$\tau_{gen}$ mass.}
		\label{fig:nnsvfit_res_rms}
	\end{center}
\end{figure}

In figure \ref{fig:nnsvfit_rescompar}, the relative differences per event between the predictions of the neural network and SVfit are shown for the different decay modes. The relative difference per event is calculated as shown in formula \ref{eq:resolution}.
\begin{equation}
\text{relative difference per event} = \frac{\mathrm{di}\mhyphen\tau_{\mathrm{gen}}\: \mathrm{mass}\;\; -\;\; \mathrm{di}\mhyphen\tau\: \mathrm{mass}}{\mathrm{di}\mhyphen\tau_{\mathrm{gen}}\: \mathrm{mass}}\:.
\label{eq:resolution}
\end{equation}
\\
The mean and the standard deviation of the histograms in figure \ref{fig:nnsvfit_rescompar} refer to the bias and the resolution of the di-$\tau$ mass reconstruction, respectively. The bias is the systematic deviation of the predictions in comparison with the target values, and is independent of the bias used in the neural network, while the resolution is a measure of the accuracy of the reconstruction. The bias in the mass determined from the neural network approach using all decay modes is -0.001, which corresponds to only -0.1 $\%$ of the mean, and is smaller than the bias using SVfit, which is 0.06 or 6 $\%$ of the mean. The mass resolution determined from the neural network using all decay modes is 0.084 while SVfit has 0.17, which is twice as large. The relative differences are also shown for all the events containing only fully leptonic $\tau$ decays, semi-leptonic $\tau$ decays, and fully hadronic $\tau$ decays. The smallest resolution is achieved for both the neural network and SVfit for events which contain the least number of neutrinos, which are the fully hadronic decays. The resolution is larger for events containing semi-leptonic decays and largest for events which contain fully leptonic decays.
\\
In figure \ref{fig:nnsvfit_res_bias}, the bias of the di-$\tau$ reconstruction and its standard error for each di-$\tau_{\mathrm{gen}}$ mass is shown for the neural network and SVfit in comparison. The neural network and SVfit have their smallest bias around 250 GeV and 100 GeV, respectively.
Figure \ref{fig:nnsvfit_res_rms} shows the comparison of the resolution per di-$\tau_{gen}$ mass using either the neural network or SVfit. The resolution of both the neural network and SVfit is constant over the mass range and is larger for SVfit as already shown in figure \ref{fig:nnsvfit_rescompar}.
\\

\subsection{Computation time of neural network and SVfit}
\label{subsec:higgsresultstime}
An important part of the comparison between the neural network and SVfit is the computation time. The benchmark SVfit algorithm requires 6 seconds of computation time for one event in our computer architecture. Our neural network algorithm requires 110 microseconds per event, a factor of fifty thousand times faster than the SVfit algorithm in the same architecture. The neural network algorithm, however, requires about 17 hours to train on 270'000 events with 400 epochs. But once the neural network is trained, the model architecture and the model weights can be saved and used later to make new predictions. In a different computing architecture, using a GPU accompanied by Tensorflow \cite{tensorflow} as the backend, the computation time to train the neural network was reduced down to 1.4 hours, with a prediction time of 7 microseconds per event. 

\subsection{Discrimination between Higgs boson Signal and Drell-Yan Background using NN and SVfit}
\label{subsec:bkganalysis}

\begin{figure}[ht]
	\begin{center}
		\includegraphics[scale= 0.6,keepaspectratio]{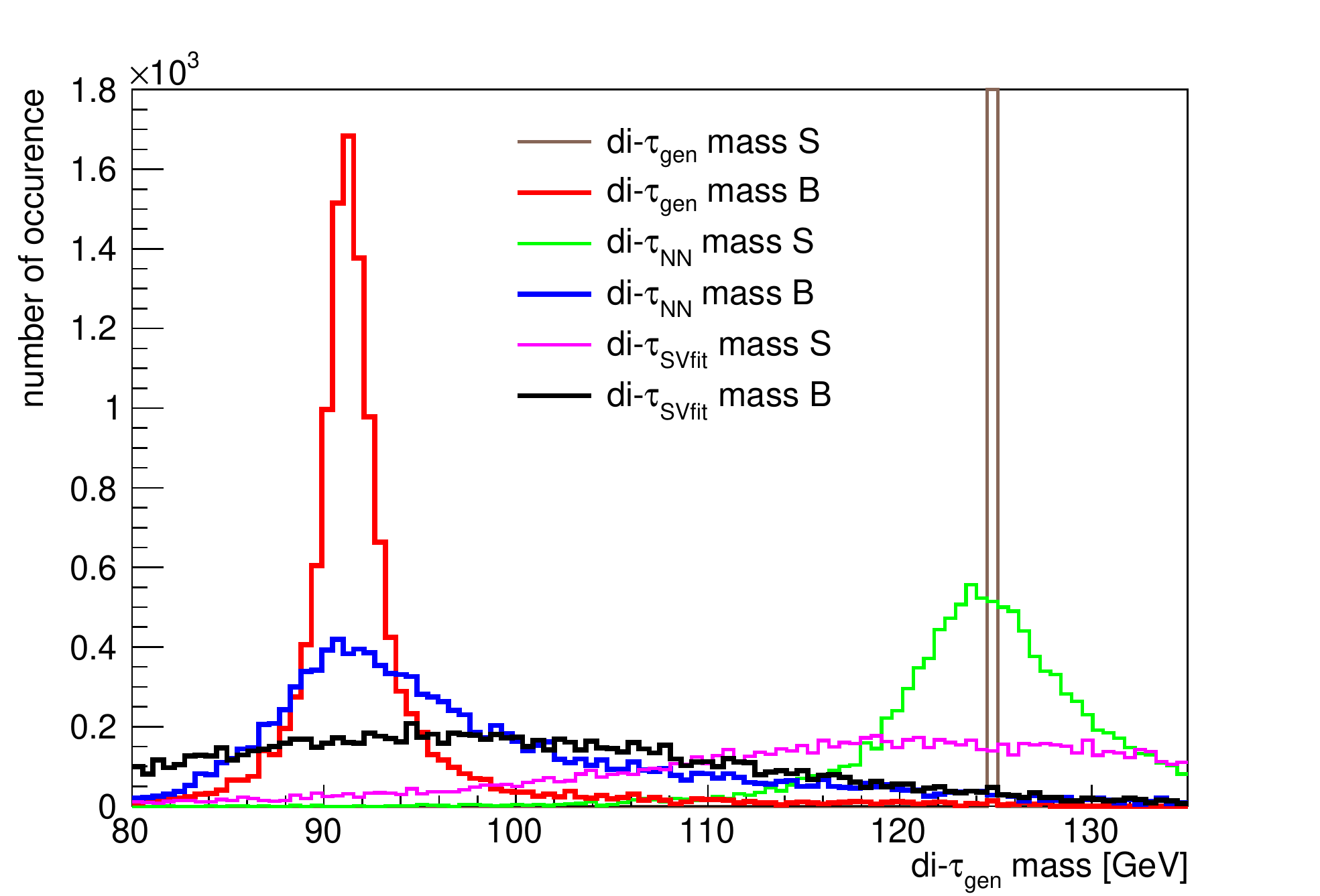} 
		\caption{Reconstructed di-$\tau$ mass for a generated Higgs boson (S) and Drell-Yan background (B). Di-$\tau_{gen}$ refers to the simulated mass values, di-$\tau_{NN}$ and di-$\tau_{SVfit}$ to the reconstructed mass values using the neural network and SVfit, respectively. }
		\label{fig:nnsvfit110GeVandDY}
	\end{center}
\end{figure}

A better discrimination between signal and background events can be achieved by an improved performance of the mass reconstruction, which can be estimated using the metric:
\begin{equation}
\text{Signal to background significance}= \frac{\epsilon_{S}\sigma_{S}\mathcal{L}}{\sqrt{\epsilon_{B}\sigma_{B}\mathcal{L}}},
\label{eq:signaltobackground}
\end{equation}
to ascertain the signal significance for finding a signal of known cross section $\sigma_{S}$ and efficiency $\epsilon_{S}$ over that of a known background, $\sigma_{B}$ and $\epsilon_{B}$, for a given integrated luminosity, $\mathcal{L}$. Figure \ref{fig:nnsvfit110GeVandDY} shows the mass reconstruction for a Higgs boson mass of 125~GeV, consistent with the CMS and ATLAS observations \cite{cmshiggsdiscovery}\cite{atlashiggsdiscovery}, also shown with the DY background peaking at the Z boson mass. The Higgs boson signal and the DY background have the same number of events and are not normalized according to their cross section. The signal to background significance in the range of 115--135 GeV for SVfit is $11.2 \pm 0.1$ and for the neural network $16.5 \pm 0.2$, where the cross section taken for the DY background is $\sigma_{B} = 1418 \pm 1\:\text{pb}$ and the cross section assumed for the production of a 125~GeV mass Higgs boson is $\sigma_{S}=0.7002 \pm 0.0006\:\text{pb}$ for a center-of-mass energy of 13 TeV, both as obtained with MadGraph~\cite{madgraph}. The integrated luminosity is set to a value of $\mathcal{L}=100\:\text{fb}^{-1}$, similar to the data sets collected by the ATLAS and CMS experiments until the end of 2017. The efficiency $\epsilon$ is the number of events in the signal range divided by the total number of events passing the requirements in section~\ref{sec:simulated_events}. The DY background events are slightly biased towards higher masses since the neural network is only trained for masses of 80~GeV and above. Using a larger mass range of the training sample would lead to a smaller deviation for the DY background and therefore to an even larger signal to background significance.

\subsection{Discussion of neural network and SVfit}
In section \ref{subsec:higgsresultsstand}, it can be seen that the reconstruction using a neural network shows a significantly smaller bias and better resolution in comparison with SVfit. The computation time, once the neural network is trained, is much shorter than using SVfit, as shown in section \ref{subsec:higgsresultstime}. Despite those advantages, the neural network has a few drawbacks. The construction of a neural network model, which delivers predictions close to the target values, is time consuming because a slight modification in the model can change the predictions drastically. This is due to the fact that the neural network is trained multiple times with a large sample of events. Unlike SVfit, a large sample of simulated events is required for training the neural network.\\
Using a neural network for reconstruction has a lot of potential, which warrants further study. A neural network is, in general, not restricted to one target parameter as used here. For example, the whole 4-vector of the di-$\tau$ system could be reconstructed if $p_{\mathrm{T}}$, $\eta$, $\phi$ and the mass are set as target parameters. If the target parameters vary over different scales, like $p_{\mathrm{T}}$ as compared to $\eta$, finding a model, and especially an activation function, which delivers predictions close to the target values for all the target parameters, is more difficult than for just one target parameter.

\section{Conclusion}
\label{sec:conclusion}
We presented a technique for the reconstruction of the mass of a di-$\tau$ system using a neural network. The predictions of the neural network are compared to the results using the SVfit likelihood technique, which is currently used by the CMS Collaboration to reconstruct the di-$\tau$ mass. For the neural network, the bias in the reconstructed mass is -0.1 \% and the mass resolution is 8.4 \%, whereas SVfit has a bias of 6 \% and a mass resolution of 17 \%. The signal to background significance of a Higgs boson signal with a mass of 125 GeV and Drell-Yan background using the neural network is $16.5 \pm 0.2$ and shows an improvement with respect to SVfit, which has a signal to background significance of 11.2 $\pm$ 0.1. Using a neural network instead of SVfit improves the performance of the di-$\tau$ mass reconstruction significantly. The neural network predicts the di-$\tau$ mass approximately fifty thousand times faster than SVfit. With the neural network technique, one must factor a few hours of retraining each time the detector reconstruction or data conditions change in order to ensure that it performs without bias. This extra computing time is not significant, and is commonplace in other machine learning software algorithms in today's experiments. Using a carefully-optimized neural network for reconstruction of the invariant mass of di-$\tau$ resonances has significant advantages over common approaches, delivering better mass resolution, reduced bias, and a much faster computation time.

\section*{Acknowledgments}
We wish to thank the Swiss National Science Foundation and the University of Zürich for their support.

\bibliography{bibfile}

\end{document}